%% file: main.tex
\documentclass[11pt, onecolumn]{IEEEtran}
\usepackage{lipsum} 
\usepackage{setspace}
\usepackage[ top    = 1in, bottom = 1in, left   = 1.25in, right  = 1.25in]{geometry}
\usepackage[T1]{fontenc}
\usepackage{graphicx}
\usepackage{cite}
\usepackage{caption}
\usepackage{subcaption}
\usepackage{epstopdf}
\usepackage{algorithm}
\usepackage{algpseudocode}
\usepackage{overpic}
\usepackage{amssymb}
\usepackage{amsmath}
\usepackage{amsthm}
\usepackage{array}
\usepackage{color}
\usepackage{url}
\usepackage{stfloats}
\usepackage{xspace}
\usepackage{etoolbox}
\usepackage{xcolor}

\IEEEoverridecommandlockouts

\theoremstyle{plain}

\newcommand{\vect}[1]{\mathbf{#1}}

\def\Htran{\mbox{\tiny $\mathrm{H}$}}
\def\Ttran{\mbox{\tiny $\mathrm{T}$}}
\def\CN{\mathcal{N}_{\mathbb{C}}} 
\def\imagunit{\mathsf{j}} 

\makeatletter 
\pretocmd\@bibitem{\color{black}\csname keycolor#1\endcsname}{}{\fail}
\newcommand\citecolor[1]{\@namedef{keycolor#1}{\color{blue}}}
\makeatother
\begin{document}
\title{Spatial Frequencies and Degrees of Freedom: 
Their Roles in Near-Field Communications}
\author{
Alva Kosasih, \"Ozlem Tu\u{g}fe Demir, Nikolaos Kolomvakis, and Emil Bj{\"o}rnson\\
\thanks{A. Kosasih, N. Kolomvakis, and E. Bj{\"o}rnson are with the Division of Communication Systems,  KTH Royal Institute of Technology, Stockholm, Sweden. (E-mail: \{kosasih,nikkol,emilbjo\}@kth.se). 
\"O. T. Demir is with the Department of Electrical-Electronics Engineering, TOBB ETU, Ankara, T\"urkiye. (E-mail: ozlemtugfedemir@etu.edu.tr).  This paper was supported by the FFL18-0277 grant from the Swedish Foundation for Strategic Research and the Grant 2022-04222 from the Swedish Research Council. \"O. T. Demir was supported by 2232-B International Fellowship for Early Stage Researchers Programme funded by the Scientific and Technological Research Council of T\"urkiye.}}
\maketitle
\begin{abstract}
As wireless technology begins to utilize physically larger arrays and/or higher frequencies, the transmitter and receiver will reside in each other's radiative near field. This fact gives rise to unusual propagation phenomena such as spherical wavefronts and beamfocusing, creating the impression that new spatial dimensions---called degrees-of-freedom (DoF)---can be exploited in the near field. However, this is a fallacy because the theoretically maximum DoF are already achievable in the far field. This paper sheds light on these issues by providing a tutorial on spatial frequencies, which are the fundamental components of wireless channels, and by explaining their role in characterizing the DoF in the near and far fields. In particular, we demonstrate how a single propagation path utilizes one spatial frequency in the far field and an interval of spatial frequencies in the near field. We explain how the array geometry determines the number of distinguishable spatial frequency bins and, thereby, the spatial DoF. We also describe how to model near-field multipath channels and their spatial correlation matrices. Finally, we discuss the research challenges and future directions in this field. 
\end{abstract}
\section{Introduction}

\label{S_Intro}

The fifth-generation (5G) cellular networks have made multiple-input multiple-output (MIMO) a mainstream technology. A typical 5G base station (BS) in the sub-6 GHz bands has hundreds of antenna elements and 32-64 transceiver chains, while mmWave bands use similar antenna numbers but fewer transceiver chains. The success of MIMO leads the way toward utilizing even larger arrays in the next-generation networks. We will call these \emph{extremely large aperture arrays (ELAA)} and note that they can either grow in physical size or be deployed at higher frequencies where the wavelength is smaller \cite{2019_Rappaport_Access}. As the array's aperture expands relative to the wavelength, the far-field distance boundary becomes so large that future wireless systems will operate predominantly in the radiative near field \cite{2020_Björnson_JCommSoc}. The channels must then be modeled differently since the wavefronts' spherical curvatures become noticeable. When this distance-dependent property is combined with MIMO, we can control the wave propagation in new ways, such as focusing signals in lens-like ways \cite{2024_Kosasih_TWC}.
This can be a paradigm shift for spatial multiplexing because the BS can simultaneously communicate with users that are either separable in the angular domain (as traditionally in the far field) or in the distance domain \cite{2023_Ramezani_Bits}. There is quantitative evidence showing that the vastly higher communication capacity can be achieved when considering near-field propagation effects \cite{2023_Bacci_WCL}. This makes it tempting to conclude that near-field systems have access to \emph{new} spatial dimensions that traditional far-field system could not exploit, but that would be a fallacy. The truth is that both systems use the same spatial dimensions but in different ways. 

In this tutorial paper, we explain how near-field channels can enable new communication features without introducing additional dimensions, thereby addressing the fallacy. {We focus on the availability of spatial dimensions from a base station's perspective, independent of use cases such as single-user or multi-user transmissions.} A key concept is the understanding of spatial frequencies and their role in wireless channel modeling. Although the same spatial dimensions are used in both near-field and far-field scenarios, they are utilized differently due to the unique characteristics of the respective scenarios. In particular, each near-field propagation path introduces  a range of spatial frequencies to the channel, while each far-field path contributes only a single one. The more spatial frequencies present in a MIMO channel, the more signals can be spatially multiplexed (i.e., the channel rank is larger), and the higher the communication capacity becomes. Hence, near-field channels are more likely to provide high capacity. Nevertheless, there are both near-field and far-field MIMO channels that can utilize all spatial frequencies and thereby enable the simultaneous transmission/reception of the same theoretically maximum number of signals. The maximum value is called the \emph{spatial degrees-of-freedom} (DoF) and depends on the array geometry. To explain this in detail, we first describe spatial frequencies in the traditional context of line-of-sight (LoS) far-field channels and present the precise connection to discrete Fourier transform (DFT) beams and the spatial DoF. We then cover recent results that extend these concepts to near-field channels. We first consider LoS and then non-LoS scenarios with multipaths. The paper ends with a review of current research challenges on this topic. Note that although this paper provides several communication examples, the fundamental near-field channel properties are also relevant for wireless localization and sensing.

\input{02_System_Model}
\input{03_Far_field}

\input{04_Near_field.tex}

\input{05_ChannelEstim}
\input{06_FutureDirections.tex}

\section{Conclusions}
\label{S_conclude}

We have provided a tutorial on how the spatial frequencies characterize multi-antenna channels. We showed that the spatial DoF are determined by the number of half-wavelength-spaced antennas and are the same for both near- and far-field channels. Far-field free-space channels utilize a single spatial frequency determined by the direction of the transmitted signal, while near-field channels are richer in the sense of utilizing multiple spatial frequencies, depending on both on the angle and distance between the user and the array. This happens because a planar wavefront in the far field corresponds to a single spatial frequency, whereas a spherical wavefront in the near field is made of a continuum of planar wavefronts. Nevertheless, we showed that the spatial frequency support is limited: it is centered around the LoS direction and the interval width decreases as distance increases or as transmission approaches the array's end-fire direction. We then provided a statistical characterization of stationary near-field multipath channels and demonstrated how the angular distribution of the scattering objects determines the DoF. Only when the angular support is limited can near-field channels achieve higher DoF than far-field channels.
We finally discussed the research challenges and future directions.

\bibliographystyle{IEEEtran}
\bibliography{IEEEabrv,myBib}
\end{document}

%% file: 02_System_Model.tex
\section{Far-field and Near-field System Model}
\label{S_mod}

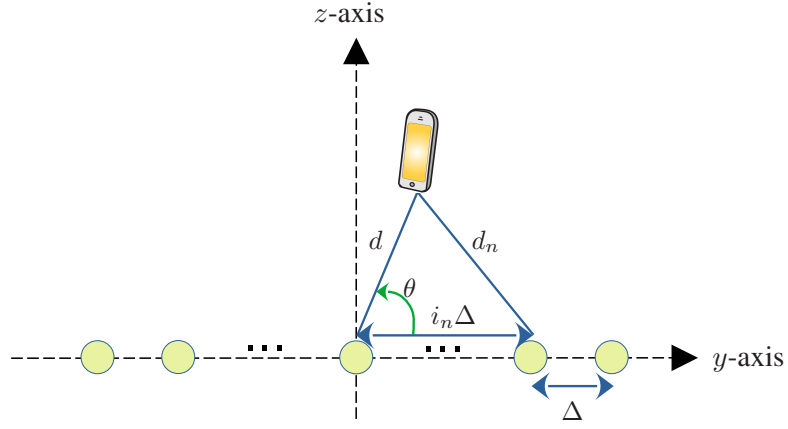
\begin{figure}
    \centering
     \begin{overpic}[width=0.6\textwidth,tics=10]{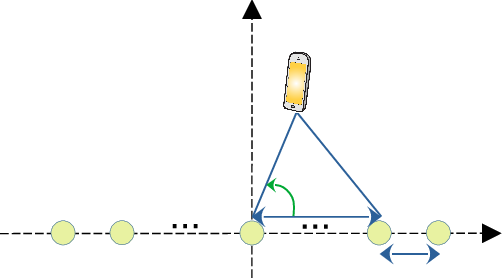}
     \put(57,18){\small $\theta$}
     \put(80,0){\small $\Delta$}
     \put(61,14){\small $i_n\Delta$}
     \put(52,25){\small $d$}
     \put(67,25){\small $d_n$}
     \put(44,58){$z$-axis}
     \put(102,8){$y$-axis}
     \end{overpic}
    \caption{The considered coordinate system where the ELAA is deployed as a ULA on the $y$-axis.}
    \label{F_ULA_illustration}
    \vspace{-3mm}
\end{figure}

We consider a user communicating with a BS equipped with an ELAA. The user is equipped with a single isotropic antenna, while the ELAA is deployed as a uniform linear array (ULA) with $N$ isotropic antennas, each connected to a transceiver chain. We assume a coordinate system where the ULA is on the $y$-axis as illustrated in Fig. \ref{F_ULA_illustration}. The $n$-th antenna is located at $(0, i_n \Delta,0 )$, where $i_n =n- \frac{N+1}{2}$, for $n=1,\ldots,N$. The spacing between adjacent antennas is denoted by $\Delta$. If $N$ is odd, then the center antenna is located at the origin: $(0,0,0)$. 
The user may be located in the reactive, radiative, or far fields of the ELAA, depending on the propagation distance:
\begin{itemize}
    \item  {The reactive near-field region of an ELAA is typically very close to the array itself and contains evanescent fields with electric and magnetic components that are out of phase, which causes energy to be stored and not radiated outwards. There are both both phase and amplitude variations across the antennas.}  
    \item The radiative near field (Fresnel region) begins and is characterized by radiated fields with spherical phase and amplitude variations among the antennas. The amplitude variations are negligible at distances greater than $d_{\rm B} = 2 D_{\rm array}$\cite{2020_Björnson_JCommSoc}, while the phase variations remain. The aperture length of the array is $D_{\rm array}=N\Delta$ for the considered ELAA.
    \item The far-field region starts after the Fraunhofer distance $    d_{\mathrm{FA}} = \frac{2D_{\rm array}^2}{\lambda}$, after which the wavefronts can be approximated as planar wavefronts with a maximum spherical-induced phase-shift variation of  $\pi/8$ across the antennas \cite{2024_Kosasih_TWC}.
\end{itemize}
The Fraunhofer distance increases quadratically when the aperture length is extended. For instance, consider a ULA with $N=10$ antennas used in MIMO systems operating at $f_c=3$\,GHz. The resulting array aperture length is $D_{\rm array}= N\cdot\frac{\lambda}{2} = 0.5$\,m, leading to a Fraunhofer distance of $d_{\mathrm{FA}} = N^2 \cdot \frac{\lambda}{2}= 5$\,m. 
 If we increase the number of antennas in the array to $N=225$ (an ELAA case) and consider a higher operating frequency of $f_c=15$\,GHz, the Fraunhofer distance significantly increases to over $500$\,m.\footnote{The 3 GHz frequency represents a typical 5G system, while the 15 GHz is a candidate band in the upper mid-band (7-24 GHz) considered for 6G because it offers a good balance between bandwidth availability and coverage \cite{2024_Bjornson_Arxiv}.}
Hence, while classical BSs only interact with users located in the far field, a future BS deployed as an ELAA will likely have a significant number of users in its radiative near field. 
In this paper, we consider the large outer part of the radiated near field where the amplitude is constant over the ELAA, but the phase varies non-linearly, and we simply refer to this as ``the near field''.

In a free-space propagation scenario, the channel response between the $n$-th antenna in the ULA and a user located at the coordinate $(\bar{x},\bar{y},\bar{z})$ can be expressed as 
\begin{equation}\label{eq:ch_resp}
 h_n = \sqrt{\beta_n} e^{-\imagunit \frac{2\pi}{\lambda} d_n },
\end{equation}
where $   d_n = \sqrt{\bar{x}^2+(i_n\Delta-\bar{y})^2+\bar{z}^2}$ is the distance between the $n$-th antenna in the ULA and a user, and  $\beta_n = \frac{\lambda^2}{(4\pi d_n)^2} \in [0,1]$ denotes the corresponding channel gain. The distance can be expressed by using the triangular cosine rule as $d_n = \sqrt{d^2 + (i_n \Delta )^2 - 2 d i_n \Delta  \cos(\theta)}$,
where $d $ is the distance of the user to the center of the array and $\theta$ is the angle between the user and the center of the array, as illustrated in Fig. \ref{F_ULA_illustration}. If $d>d_{\rm B}=2 D_{\rm array}$, then the channel gain is approximately constant among the antennas \cite{2020_Björnson_JCommSoc} so that 
$\beta_n=\beta = \frac{\lambda^2}{(4\pi d)^2}$, $\forall n$ 
\cite{bjornson2024towards}. If we collect the channel responses in a vector, we can write it as
\begin{equation}
    \vect{h} = [h_1, \ldots, h_N] ^{\Ttran}= h_c  \vect{b} (\theta,d),
\end{equation}
where $\vect{b}(\theta,d) = \left[ e^{- \imagunit \frac{ 2  \pi (d_1- d) }{\lambda}  } , \dots,  e^{- \imagunit\frac{ 2  \pi (d_N - d) }{\lambda}  }  \right]^{\Ttran}$ is called the \emph{near-field array response vector} and $h_c \triangleq \sqrt{\beta}  e^{\frac{-\imagunit 2 \pi d}{\lambda}}$ is a scalar.
The distance can also be expressed as $d_n = d\sqrt{1 + \frac{(i_n \Delta )^2}{d^2} - \frac{2 i_n \Delta}{d} \cos(\theta)}$. If  we apply the second-order Taylor approximation $\sqrt{1+x}\approx 1+\frac{1}{2}x-\frac{1}{8}x^2$ (that is accurate for small $x$), we obtain the \emph{near-field expansion} \cite{ziomek1993three}
\begin{align}  \label{eq:near-field-expansion}
&d_{n}\approx  d -\Delta\Big(i_n\cos(\theta)\Big)+\Delta^2\left(\frac{i^2_n-i_n^2\cos^2(\theta)}{2d}\right).
\end{align}
When doing so, we disregard the terms involving $1/d^2, \ldots, 1/d^4$ as they are negligible in the radiative near-field region. The last term in \eqref{eq:near-field-expansion} depends non-linearly on the antenna index $i_n$, which shows that the channel vector describes a non-planar wave. 
When $d$ exceeds the Fraunhofer distance, the last term can also be neglected. In this case, $\vect{b}(\theta,d) 
\approx \left[ e^{\imagunit\frac{2\pi}{\lambda}i_1\Delta\cos(\theta)}, \ldots, 
    e^{\imagunit\frac{2\pi}{\lambda}i_N\Delta\cos(\theta)}
    \right]^{\Ttran} = \vect{a}(\theta)$, which is the classic expression for a \emph{far-field array response vector}.
This expression is independent of the distance $d$, so we can only resolve the angle. 

%% file: 03_Far_field.tex
\section{Spatial Frequencies and Degrees-of-Freedom in the Far-Field Region}

In this section, we characterize far-field channels using spatial frequencies, orthogonal beam grids, and spatial DoF. Each spatial frequency corresponds to a spatial resource that can be used to direct a beam in a specific direction. 
Considering an ELAA with $N$ antennas, we can create a grid of, at most, $N$ orthogonal beams by sampling the spatial frequencies. The spatial DoF is the maximum number of orthogonal beams that can be generated for a given array. Hence, the spatial DoF determines the maximum number of spatial layers that can be simultaneously transmitted/received using it  \cite{bjornson2024towards}.
{The receiving users' locations, antenna configurations, and mutual channel characteristics determine how many spatial layers can be utilized at any given time. One extreme case is a single-user far-field MIMO LoS scenario, where only one layer is utilized among the $N$ available spatial frequency resources. The other extreme is a setup with $K=N$ users, well separated in the angular domain, allowing all $N$ spatial frequency resources to be utilized. We will elaborate further on this in this section.}

\subsection{Spatial Frequencies}

When a time-domain signal is observed over time at a fixed location, it will oscillate based on what \emph{temporal frequencies} it contains. For example, the complex exponential signal $e^{\imagunit 2\pi  f_c t}$ has a temporal frequency of $f_c$ since the time interval between two identical values is $\frac{1}{f_c}$. 
Suppose the user in Fig.~\ref{F_ULA_illustration} is in the far field and transmits an electromagnetic wave. The channel coefficient between the user and the point $(0,y,0)$ in the ELAA is $h_ce^{\imagunit\frac{2\pi}{\lambda}y\cos(\theta)}$.
Therefore, the signal observed at this point is $h_ce^{\imagunit2\pi f_c t}e^{\imagunit\frac{2\pi}{\lambda}y\cos(\theta)}$, corresponding to the multiplication of the emitted signal and the channel. 
Considering an arbitrary fixed time instance (e.g., $t=0$), we can determine the \emph{spatial frequency} of the observed signal: the complex exponential repeats itself after a distance  $ \frac{\lambda}{|\cos(\theta)|}$ along the $y$-axis.
The exact spatial frequency is $\cos(\theta)/\lambda$, and because $\cos(\theta) \in [-1,1]$, it can be any value in the interval $\left[-\frac{1}{\lambda}, \frac{1}{\lambda}\right]$.
A transmitting user in the positive/right quadrant in Fig.~\ref{F_ULA_illustration} gives a positive spatial frequency, while a user in the negative/left quadrant gives a negative value. 
The array observes the spatial frequency, $\pm \frac{1}{\lambda}$, 
when the signal propagates along the $y$-axis (i.e., $\theta=0$ or $\theta=\pi$). Conversely, the array observes zero spatial frequency when the signal propagates perpendicular to the $y$-axis (i.e., $\theta=\pi/2$).
In a nutshell, an array can estimate the angle-of-arrival by measuring the spatial frequency content of the impinging wavefront.

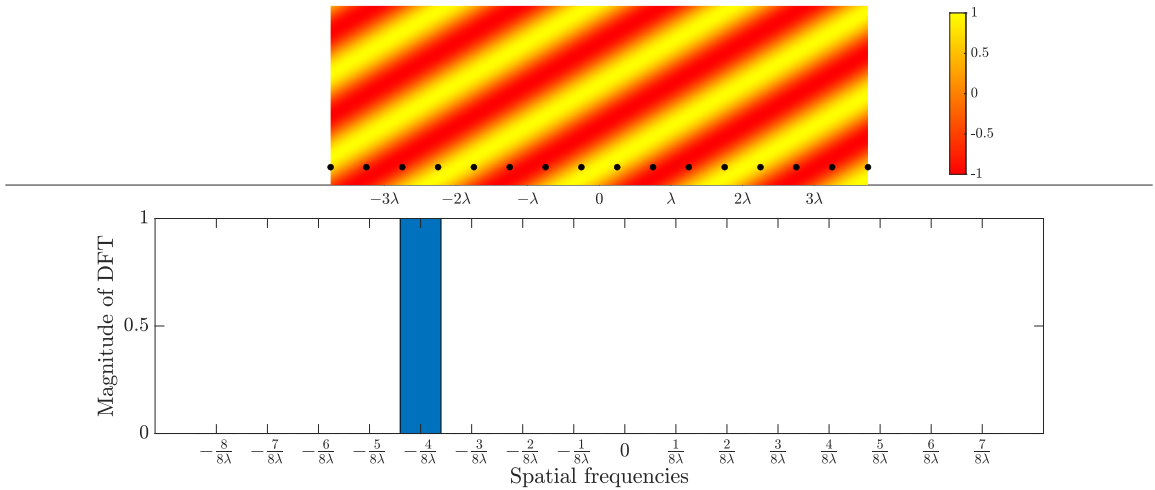
\begin{figure}
    \centering
    \begin{overpic}[width=\textwidth,tics=10]{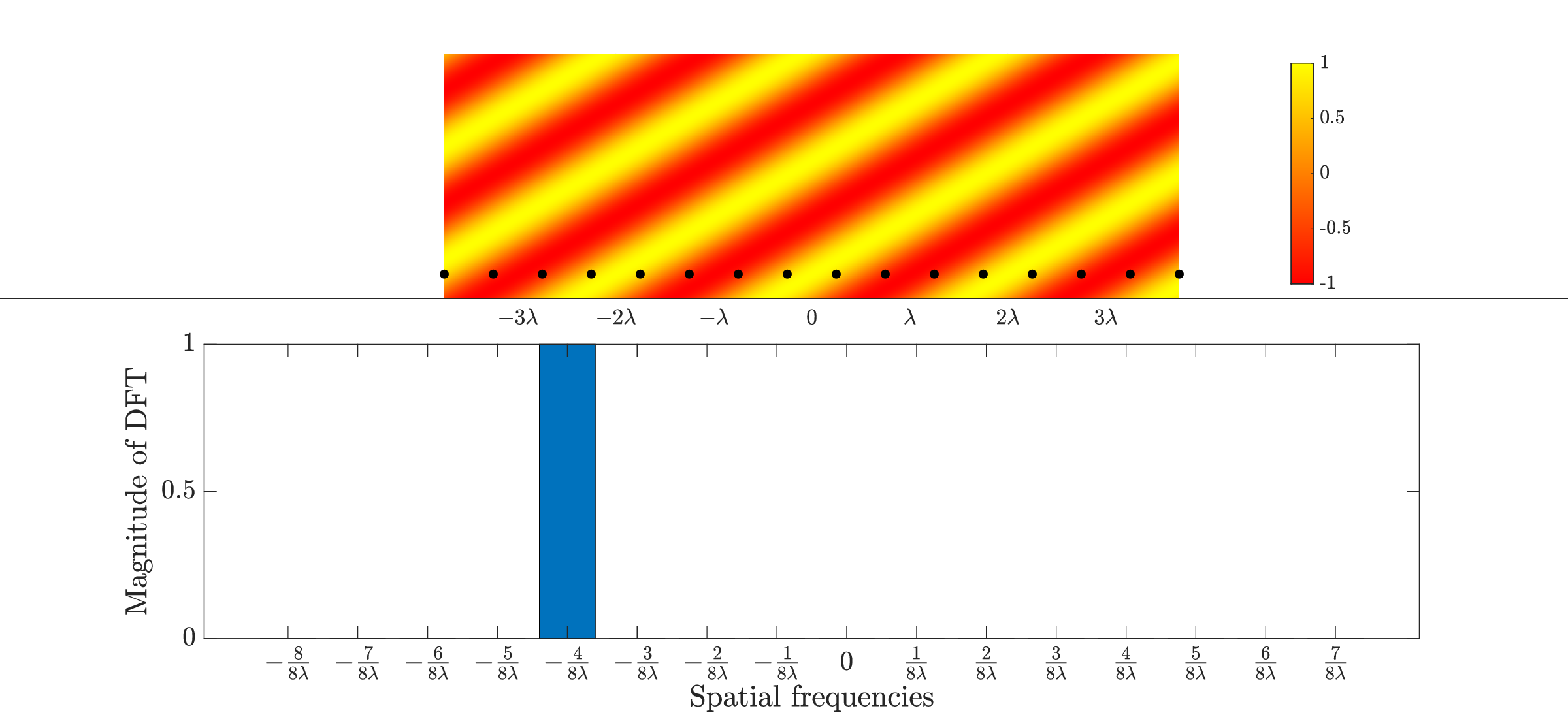}
    \put(19,30){\vector(0,1){5}}
    \put(19,30){\vector(1,0){5}}
      \put(25,29.5){\small $y$}
        \put(18.5,36){\small $z$}
\end{overpic}
\caption{The spatial variation and the magnitude of the DFT of the observed signal when the user is located in the far field of the array with $\Theta = -1/2$. The top figure displays the normalized real part of the observed signal {with values shown in the color bar}.}
    \label{fig:spatial-frequencies}
       \vspace{-3mm}
\end{figure}

In Fig.~\ref{fig:spatial-frequencies}, we assume that the user transmits from the direction $\theta = 2\pi/3$ with the directional cosine $\Theta \triangleq \cos(\theta) = -1/2$. At the top of this figure, we show the normalized real part of the observed signal in the $yz$-plane at $t=0$. There are $N=16$ antennas with locations shown as black dots, and these antennas sample the signal  $h_ce^{\imagunit2\pi f_c t}e^{\imagunit\frac{2\pi}{\lambda}y\cos(\theta)}$ along the $y$-axis. {As this figure illustrates, incoming waves behave as planar waves when they arrive from the far-field of the array.} The magnitude of the $N=16$ point DFT of the received signal is shown in the bottom part of the figure. It has a peak at the spatial frequency $\frac{\cos(\theta)}{\lambda}=-\frac{4}{8\lambda}$. Hence, we observe a single spatial frequency when a far-field signal impinges from a single direction. If the spatial frequency differs from the DFT bins, spectral leakage to adjacent bins will occur (as usual when analyzing sampled signals).

\subsection{Grid of Orthogonal Beams}

To ensure that a signal reaches a prospective user regardless of its location, we need to transmit using a collection of beams that cover all angles.
A collection/grid of beams is said to be \emph{orthogonal} if each beam's peak coincides with all other beams' nulls. 
A common way to generate such a grid is to consider so-called DFT beams, generated by sampling the spatial frequency domain $\left[-\frac{1}{\lambda}, \frac{1}{\lambda}\right]$ and computing the inverse DFT for each of these frequencies.
The DFT beam, obtained from the spatial frequency $\cos(\theta)/\lambda$, 
can be expressed using the far-field array response vector and directional cosine $\Theta = \cos(\theta)$ as
\begin{align}
    \mathbf{a}({\Theta}) = e^{\imagunit\frac{2 \pi}{\lambda}\frac{1-N}{2}\Delta \Theta}\left[ 1, e^{\imagunit \frac{2 \pi}{\lambda} \Delta {\Theta}}, \dots, e^{\imagunit \frac{2 \pi}{\lambda} (N-1)\Delta {\Theta} }\right]^{\Ttran}.
\end{align}
The range $[-1,1]$ of directional cosine values represents the range of normalized spatial frequencies.
By constructing a grid of beams through uniform sampling of the interval $[-1,1]$, we obtain beam directions
\begin{align} \label{eq:directions}
\Theta_n = \frac{\left\lfloor \frac{N}{2} \right\rfloor+1-n}{N} \cdot \frac{\lambda}{\Delta}, \quad \forall n\in\{1, \ldots, N\} \text{ so that }\Theta_n \in[-1,1].
\end{align}
Notice that the directional cosine is sampled with a period proportional to $\frac{\lambda}{\Delta N}$.
From \eqref{eq:directions}, we observe three possible cases with respect to the antenna spacing $\Delta$:

1) $\Delta=\frac{\lambda}{2}$: We obtain $N$ orthogonal beams, each corresponding to one normalized spatial frequency in the range $[-1, 1]$ since the maximum absolute value of $\Theta_n$ is $1$.

2) $\Delta>\frac{\lambda}{2}$: We obtain $N$ orthogonal beams, but the corresponding normalized spatial frequencies take values from a smaller subset of $[-1, 1]$ since the maximum absolute value of $\Theta_n$ is less than $1$.

3) $\Delta<\frac{\lambda}{2}$: We obtain fewer than $N$ orthogonal beams, resulting in a total number of spatial frequencies less than $N$. This is because beams corresponding to $|\Theta_n|>1$ are discarded as $|\cos(\theta)|$ cannot exceed one.

We conclude that we can construct a grid of orthogonal  $N$ DFT beams as long as $\Delta\geq \frac{\lambda}{2}$. However, the larger $\Delta$ becomes, the smaller the range of spatial frequencies where all the beams are located.
Nevertheless, we cannot create more than $N$ orthogonal beams because the array cannot distinguish high spatial frequencies from smaller ones, where the former is a result of aliasing. Moreover, a large antenna spacing gives spatial undersampling that leads to aliasing in the angular domain. A single beam can then point in multiple directions (called grating lobes).
On the other hand, when $\Delta$ becomes smaller than $\frac{\lambda}{2}$, the number of orthogonal beams becomes smaller than $N$.
In all cases, the number of orthogonal beams is upper bounded by $N$.

We now set $\Delta=\frac{\lambda}{2}$, which is the optimal antenna spacing that avoids aliasing. 
By aggregating the phase-shifted versions of all $N$ DFT beams into a matrix, we obtain 
\begin{align} \label{eq:DFTmatrix}
    \mathbf{A}(\vect{\Theta}) = e^{-\imagunit\frac{2 \pi}{\lambda}\frac{1-N}{2}\Delta \Theta_n}\left[ \vect{a}({\Theta}_{1}), \dots, \vect{a}({\Theta}_{n}), \dots, \vect{a}({\Theta}_{N}) \right].
\end{align}
The $n$-th column of this matrix is 
\begin{align}
   e^{-\imagunit\frac{2 \pi}{\lambda}\frac{1-N}{2}\Delta \Theta_n}\vect{a}({\Theta}_{n})=   \left[1, e^{-\imagunit 2\pi\frac{n-1-\lfloor N/2 \rfloor}{N}},  \ldots,  e^{-\imagunit 2\pi (N-1)\frac{n-1-\lfloor N/2 \rfloor}{N}} \right]^{\Ttran},
\end{align}
which is the column of the DFT matrix corresponding to the frequency bin $n-1-\left\lfloor \frac{N}{2} \right\rfloor$.
Since the DFT matrix $\mathbf{A}(\vect{\Theta})$ is a (scaled) unitary matrix, all columns are mutually orthogonal. Hence, if the user's angular direction matches with the beam direction, $\Theta=\Theta_n$, the inner product between the vector $\vect{a}({\Theta}_{n})$ and the far-field channel $\vect{a}({\Theta})$ is $N$. Conversely, it is zero if the user's angular direction corresponds to another angle  selected from the grid, $\Theta=\Theta_i$.
{
We mathematically express this in terms of the correlation between the array response vectors for different angular directions as
\begin{equation}
    \frac{|\vect{a}^{\Htran}(\Theta) \vect{a}(\Theta_n)|}{N} = 
\begin{cases} 
      1, & \text{if } \Theta = \Theta_n, \\
      0, & \text{if } \Theta = \Theta_i.
   \end{cases}
\end{equation}
}

\begin{figure}
    \centering
     \begin{overpic}[width=0.52\textwidth]{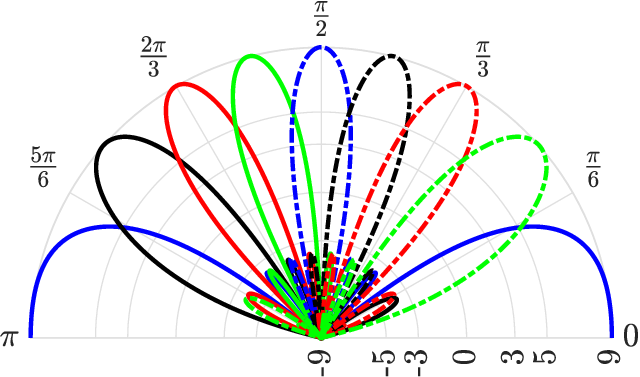}
     \put(0,55){Observation angle ($\theta$)}
     \put(-5,1){Beamforming Gain (dB)}
     \end{overpic}
    \caption{Grid of eight orthogonal beams shown using with respect to the  observation angle. {The outer-most beam is divided into two parts, one aimed at angle $0$ and another aimed at $\pi$.}}
    \label{F_grid_ortho}
       \vspace{-3mm}
\end{figure}

We further plot $|\vect{a}^{\Htran}(\Theta) \vect{a}(\Theta_n) |^2/N$ in Fig.~\ref{F_grid_ortho}, which represents the beamforming gain \cite[Ch.~4.3.3]{Bjornson_BOOK}, for the observation angles $\theta \in [0,\pi]$ and $\Theta = \cos(\theta)$. The plot shows that the orthogonal beams' peaks coincide with the others' null points.
Interestingly, the beams have different angular widths. 
{Therefore, we cannot sample the angular directions uniformly.
Instead, we must sample the directional cosines uniformly. The inverse function $\theta = \cos^{-1}(\Theta)$ is non-linear, particularly close to the end-fire directions.
This implies that the array has a lower angular resolution in these directions, manifesting as a wider beamwidth.}

\subsection{ Spatial Degrees-of-Freedom}

An ELAA can transmit and receive in $N$ spatially orthogonal angular directions simultaneously using DFT beams. In an ideal practical scenario, the $N$ users are located exactly in the directions specified by \eqref{eq:directions}, for $\Delta = \lambda/2$. The ELAA can then apply a DFT to the received signal to separate the users, with each signal occupying a unique DFT bin. The spatial DoF is $N$ and we can fully utilize it to multiplex $N$ user signals, resulting in an $N$-fold increase in capacity compared to a single-user scenario. This is known as the \emph{spatial multiplexing gain}.
In practice, the users will likely be distributed over the angles in a less regular manner, but one can still achieve a multiplexing gain equal to the number of linearly independent channel vectors. However, orthogonality must instead be created by signal processing at the ELAA, such as zero-forcing beamforming 
\cite{Bjornson_BOOK}.
If an individual user has multiple antennas, we can send several spatial layers of data to them if the respective channel vectors are linearly independent. However, this is not the case in the far-field free-space LoS scenario considered.

Note that if $\Delta \geq \frac{\lambda}{2}$, we can  construct $N$ orthogonal beams, otherwise, the number of orthogonal beams is  
$\left \lfloor2N\frac{\Delta}{\lambda} \right \rfloor< N$, implying reduced DoF. Hence, the DoF for a ULA with a spacing of $\Delta \leq \frac{\lambda}{2}$ can be expressed (approximately) as
\begin{align}
    2N\frac{\Delta}{\lambda} = \frac{2}{\lambda}D_{\rm array},
\end{align}
where $D_{\rm array}=N\Delta$ is the aperture length. This can also be proved by studying the observable electromagnetic fields \cite{franceschetti2017wave,hu2018beyond,pizzo2022nyquist}. Hence, the maximum DoF is $2/\lambda$ per meter of the ULA. Different expressions can be obtained for two-dimensional array geometries \cite{pizzo2022nyquist}.

%% file: 04_Near_field.tex
\section{Spatial frequencies and Degrees-of-Freedom in the near-field region}
\label{secso}

In the previous subsection, we concluded that it is impossible to send multiple streams to a multi-antenna user in far-field free-space communication. However, this does not hold in near-field communications because the spherical curvature and narrow beams make the channel from the different user antennas to the ELAA distinguishably different \cite{cui2022near}. 
Another unique property in near-field free-space communications is that we can serve users located in the same direction simultaneously. As long as they are at different distances, the wavefronts will have different spherical curvature that enables the ELAA to separate them by signal processing (e.g., zero-forcing) \cite{2023_Ramezani_Bits,bjornson2024towards}. 
In fact, the term ``beam'' becomes outdated because a signal transmitted to a near-field user will not look like a classical cone but be focused in an ellipsoidal region around the user with limited energy leakage outside the focal area \cite{2024_Kosasih_TWC,zhang20236g}.
This is called \emph{beamfocusing} instead of beamforming. 
Due to these novel features, the near field is often causally said to provide new dimensions or DoF.
However, as we will explain in the following, the same dimensions are utilized in a more efficient manner, particularly in LoS scenarios.

\subsection{Representation of Near-Field Wavefront Using Orthogonal Beams}

\begin{figure}
    \centering
    \begin{overpic}[width=\textwidth,tics=10]{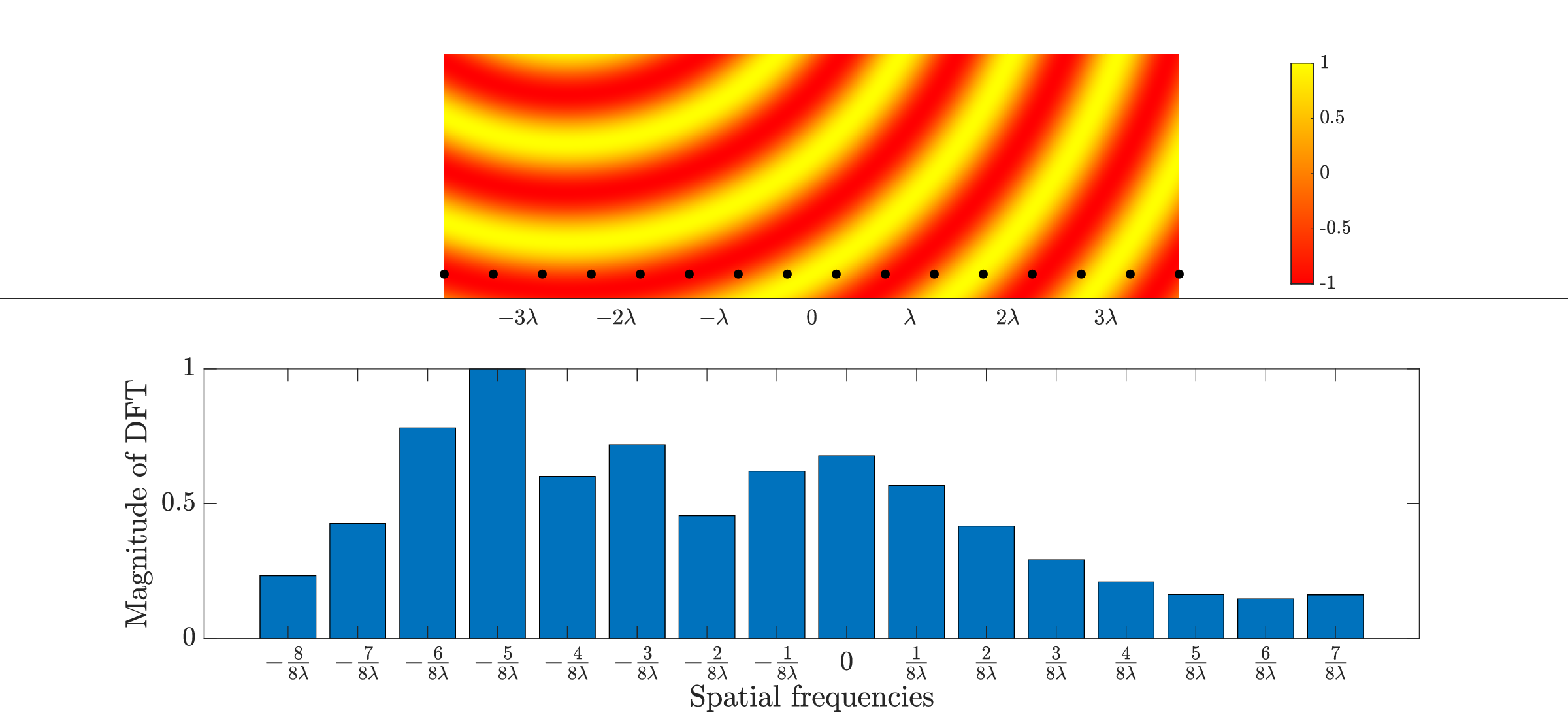}
    \put(19,30){\vector(0,1){5}}
    \put(19,30){\vector(1,0){5}}
      \put(25,29.5){\small $y$}
        \put(18.5,36){\small $z$}
\end{overpic}
\caption{The spatial variation and the magnitude of the DFT of the observed signal when the user is located in the near field of the array with $\Theta = -1/2$ and $d=5\lambda$. The top figure displays the normalized real part of the observed signal  {with values shown in the color bar}.}
    \label{fig:spatial-frequencies-near-field}
       \vspace{-3mm}
\end{figure}
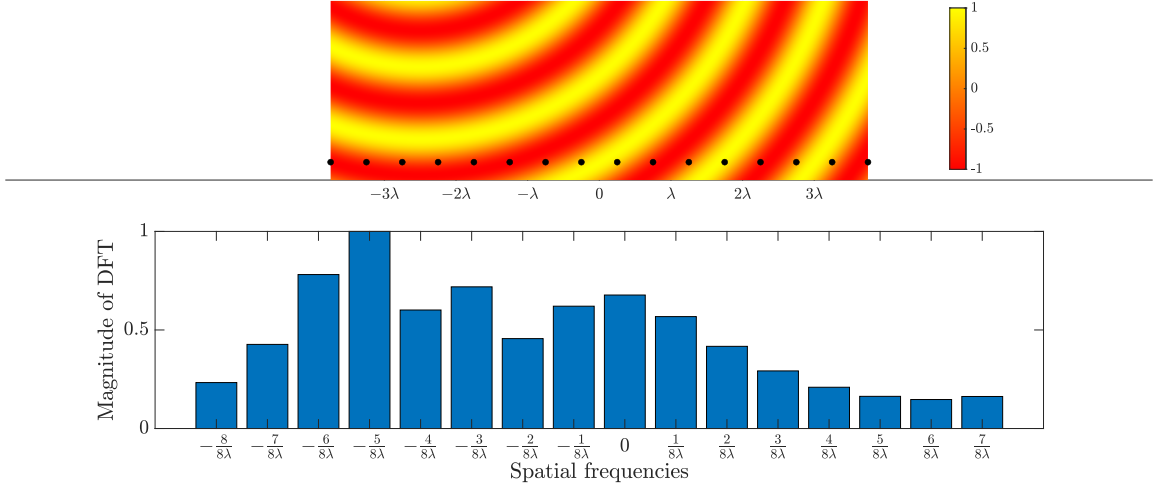

In far-field analysis, we learned that the number of distinguishable spatial frequencies is limited by the number of antennas, $N$.
This upper bound and channel characterization also hold true in the near field.  We show this by plotting  Fig.~\ref{fig:spatial-frequencies-near-field}, where we consider the same setup as in Fig.~\ref{fig:spatial-frequencies} except that the user is located in the near field of the ELAA with $\Theta= -1/2$ and $d=5\lambda$. At the top of this figure, we show the normalized real part of the observed signal in the $yz$-plane at $t=0$  {by omitting the amplitude variations across the antennas.} The magnitude of the $N=16$ point DFT of the received signal is shown in the bottom part of the figure.
Even if the signal impinges from a single LoS path, the receiver observes a wide range of spatial frequencies---instead of only $-\frac{4}{8\lambda}$ as in the far-field case.
This makes the near-field LoS channel vector ``richer'' in terms of spatial frequency content than the corresponding far-field LoS channel. Nevertheless, the range of spatial frequencies is the same as in far-field communications. 

Another way to view this is that the received signal has a planar wavefront in the far field with a single spatial frequency. In contrast, in the near field, the spherical wavefront can be expressed as a continuum of planar wavefronts according to the \emph{Weyl integral} \cite{weyl1919ausbreitung}, implying that multiple spatial frequencies will be occupied.
Although theoretically, the summation extends to infinity, Fig.~\ref{fig:spatial-frequencies-near-field} shows that   $N$ planar wavefronts contribute to forming the spherical wavefront. 
More specifically, if we take the inverse DFT of the spatial frequency spectrum, we can synthesize the sampled spherical wave as the sum of (up to) $N$ planar waves. 

We now consider a BS equipped with an ELAA (with $N$ antennas and $\Delta = \lambda/2$) that serves a single antenna user located in the near field of the BS with an angular direction of ${\theta}$ and distance $d$. Recall that the DFT matrix $\vect{A({\Theta}})$ in \eqref{eq:DFTmatrix} is a unitary matrix scaled by $\sqrt{N}$. Since $\vect{A({\Theta}})\vect{A}^{\Htran}(\vect{\Theta}) = N \vect{I}_N$, the near-field array response vector can be expressed as 
\begin{align}\label{eq:DFT_coeff_Near_Field}
 \vect{b}(\overline{\Theta},d)=  \vect{A({\Theta}})\underbrace{\frac{\vect{A}^{\Htran}(\vect{\Theta})\vect{b}(\overline{\Theta},d)}{N}}_{\text{DFT coefficients}},
\end{align}
where $\overline{\Theta} = \cos({\theta})$ is the directional cosine.
The  squared magnitude of the DFT coefficients, specifying the gain associated with each spatial frequency, can be written as
\begin{equation}\label{eq:norm_gain_DFTBeam}
    G\left(\overline{\Theta},d,{\Theta}_{n}\right) = \frac{  \left|  \vect{a}^{\Htran}({\Theta}_{n}) \vect{b}(\overline{\Theta},d)  \right|^2}{N^2},
\end{equation}
for $n=1,\ldots,N$.
Utilizing the near-field expansion in \eqref{eq:near-field-expansion}, we obtain $d_n = \sqrt{d^2 + (i_n \Delta )^2 - 2 d i_n \Delta  \overline{\Theta}} \approx d - i_n \Delta \overline{\Theta}  +  \frac{i_n^2  \Delta^2 (1-\overline{\Theta}^2) }{2d} $. Hence, we can approximate
\begin{align}\label{eq:expanded_DFT_correl}
 \frac{\left| \vect{a}^{\Htran}({\Theta}_{n}) \vect{b}(\overline{\Theta},d)  \right |}{N} &\approx \frac{1}{N}  \left| \sum_{i_n }
 e^{\imagunit \frac{2 \pi}{\lambda} \left(  i_n \Delta \overline{\Theta}  - \frac{i_n^2  \Delta^2 (1-\overline{\Theta}^2) }{2d} \right)   - 
 \imagunit \frac{2 \pi}{\lambda} i_n \Delta {\Theta}_{n} } \right| 
  = \frac{1}{N} \left| \sum_{n=1}^{N} e^{\imagunit\pi \left( c_1 n - c_2\right)^2 } \right|,
\end{align}
where we introduce the notation $c_1 = \frac{\gamma}{\sqrt{2}}  $,  $\gamma = \sqrt{\frac{\Delta (1- \overline{\Theta}^2)}{d} }$,
and $c_2 =  \frac{1}{2 c_1} \left(  \overline{\Theta} - {\Theta}_{n}+  (N+1) c_1^2    \right)$.
When $N$ is large, as intended for ELAAs, we can compute an accurate closed-form expression for \eqref{eq:expanded_DFT_correl} by replacing the summation with an integral. More specifically, this approximation is expressed as  \cite{2023_CYou_CommLett}
\begin{align}
   \frac{|  \vect{a}^{\Htran}({\Theta}_{n}) \vect{b}(\overline{\Theta},d)  |}{N} 
   {\approx}& \left| \frac{ C(\beta_1 + \beta_2  )  +   C(\beta_1 - \beta_2)  + \imagunit \left(   S(\beta_1 + \beta_2 ) + S(\beta_1 - \beta_2)   \right) }{2\beta_1} \right|,
\end{align}
where $\beta_1 = \frac{N}{2} \sqrt{\frac{\Delta (1- \overline{\Theta}^2)}{d} }  $, $\beta_2 = \frac{1}{\gamma}(\overline{\Theta}- {\Theta}_{n})$, and
the functions $C(\cdot )$ and $S(\cdot )$ are the Fresnel integrals\footnote{Both Fresnel integrals are odd functions: $C(-v) = -C(v)$ and $S(-v) = -S(v)$.} \cite{1956_Polk_TAP}. The approximation  is accurate when we have many antennas.
Therefore, we can obtain a closed-form solution that approximates the gain function in \eqref{eq:norm_gain_DFTBeam} as
\begin{equation}\label{eq:closed_form_gain_DFTBeam}
     G(\overline{\Theta},d,\Theta_n) \approx  \frac{\left[ C(\beta_1 + \beta_2) +   C(\beta_1 - \beta_2)\right]^2 + \left[ S(\beta_1 + \beta_2) +   S(\beta_1 - \beta_2)\right]^2}
      {4 \beta_1^2 }.
\end{equation}
For a user at a given location ($\overline{\Theta},d$), $\beta_1$ becomes a constant and the gain function only depends on the variable ${\Theta}_{n}$, which appears in the expression of $\beta_2$.  The resulting gain function allows us to analyze the presence of the $n$-th spatial frequency $\Theta_n$ in the considered user's channel. We notice that the gain function is symmetric with respect to $\beta_2$, in the sense that $ {G}(\beta_2) =  {G}(-\beta_2)$ \cite{2024_CYou_TWC}. 
 Hence, for each value of $\overline{\Theta}$, there exists a unique range of spatial frequencies (${\Theta}_{n}, n=1,\dots,N$), centered at $\overline{\Theta}$, where the DFT coefficients (gain) are significant.

\begin{figure*}
\centering
\subfloat[$\theta = \frac{\pi}{2}$]
{\includegraphics[width=0.5\textwidth]{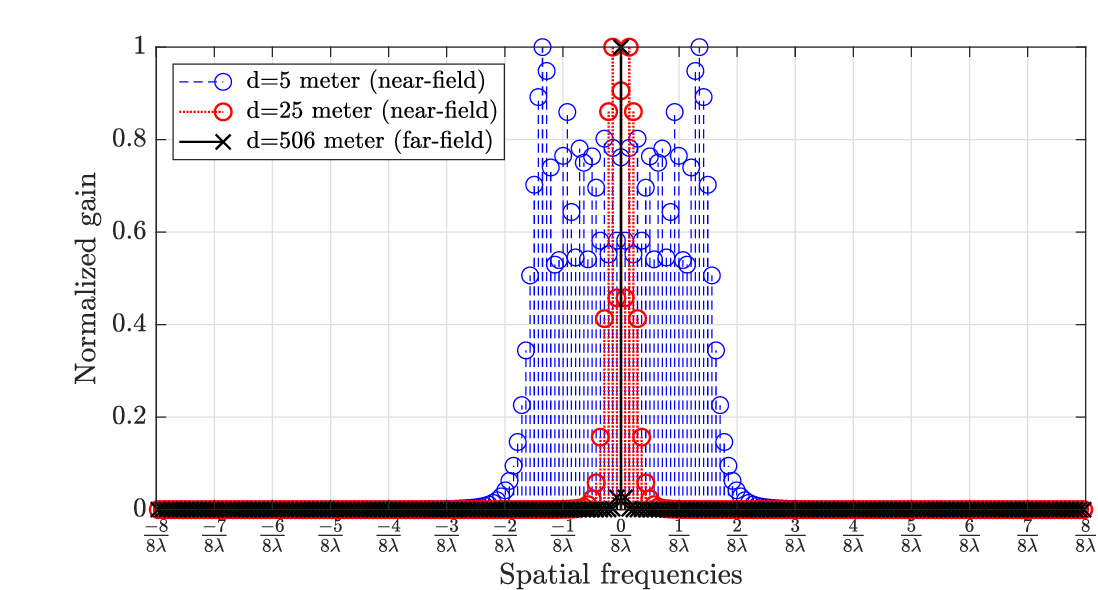}}\hfill
\centering
\subfloat[$\theta = \frac{\pi}{3}$]
{\includegraphics[width=0.468\textwidth]{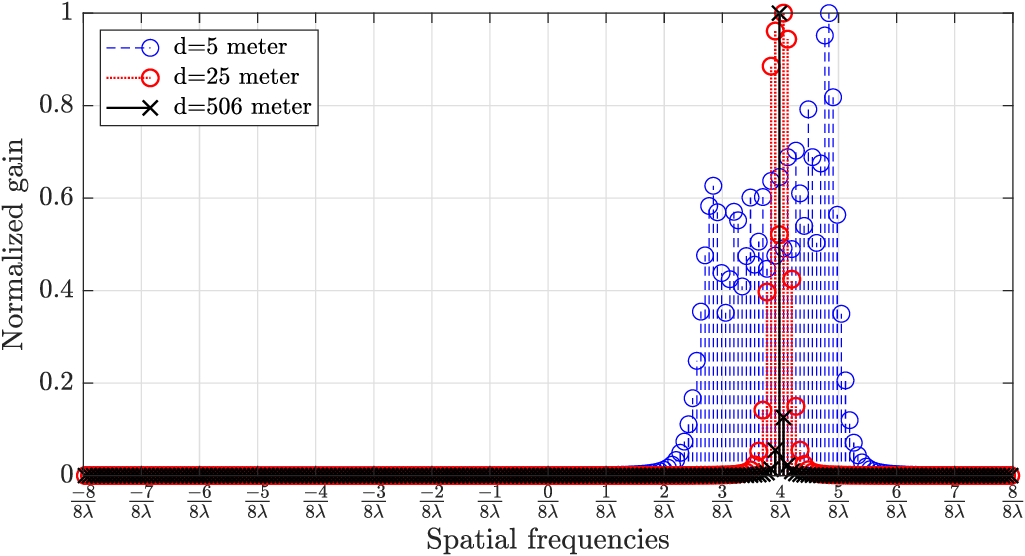}}
\caption{The normalized gain for every spatial frequency for a user located at $(\theta,d)$. We consider a carrier frequency of  {$15$}\,GHz and $N=225$ antennas with half-wavelength spacing. }
\label{F_gain_func}
   \vspace{-3mm}
\end{figure*}

 In Fig.~\ref{F_gain_func}, we consider $N=225$, and a carrier frequency of $f_c=15$\,GHz, corresponding to the Fraunhofer distance of  $d_{\rm FA} = 2 D_{\rm array}^2/\lambda  {\approx 506}$\,m. We plot the normalized gain of the spatial frequencies for one far-field distance ($d=506$\,m) and two near-field distances ({$d=5$}\,m and $d=25$\,m). In the case of far-field case, 
 there is only one non-zero power (indicated by the black curve) at the spatial frequency $\frac{\cos(\pi/2)}{\lambda}=0$ in Fig.~\ref{F_gain_func}(a) and $\frac{\cos(\pi/3)}{\lambda}=\frac{1}{2\lambda}$ in Fig.~\ref{F_gain_func}(b). 
 The blue curves in Fig.~\ref{F_gain_func} depict the closest near-field case of  {$d=5$}\,m. 
 We notice that the spatial frequencies with non-zero gains form a window centered at $\overline{\Theta}$, aligning with the theoretical analysis above. This applies to both the broadside and non-broadside scenarios considered in Fig.~\ref{F_gain_func}(a) and \ref{F_gain_func}(b), respectively.
In the motivational example in Fig.~\ref{fig:spatial-frequencies-near-field},
the near-field channel contained all spatial frequencies. The major difference in this example is that we have many more antennas, in which we can clearly see that the channel has a limited spatial bandwidth, even if it is much larger than in the far field.
 The range of used spatial frequencies can be searched based on the analytical gain function in \eqref{eq:closed_form_gain_DFTBeam}.

\begin{figure*}
\centering
\subfloat[Varying distance ($d$)]
{\includegraphics[width=0.5\textwidth]{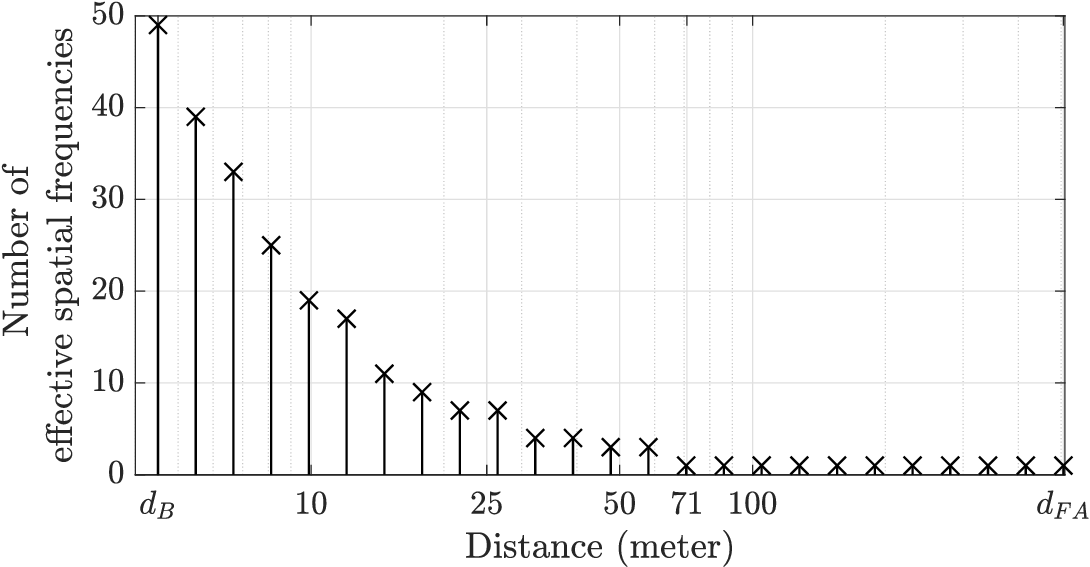}}\hfill
\centering
\subfloat[Varying angle ($\theta$)]
{\includegraphics[width=0.5\textwidth]{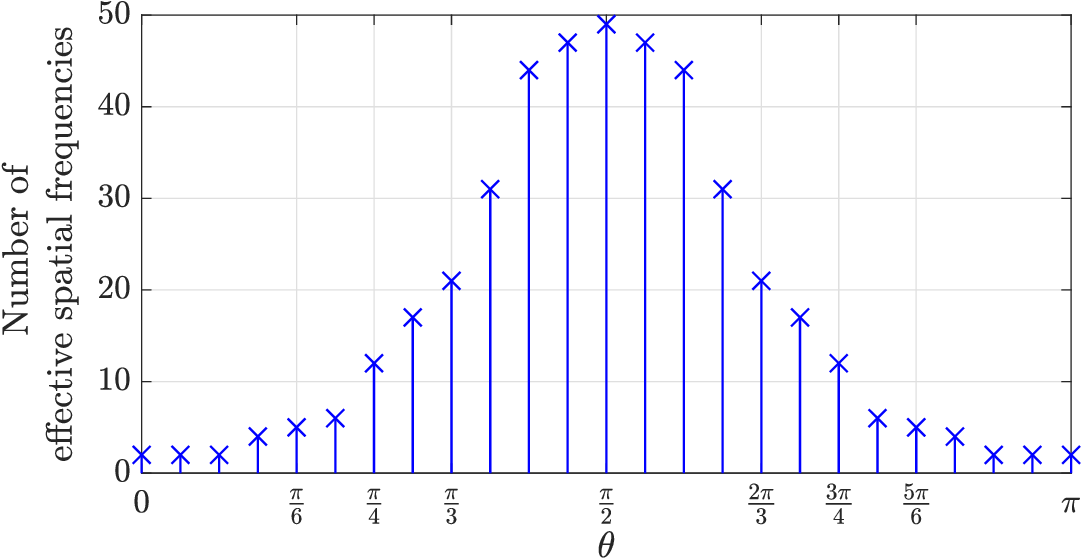}}
\caption{The number of effective spatial frequencies with respect to distance and angle, $N=225$ and $f_c= 15$\,GHz. }
\label{F_eff_spat}
   \vspace{-3mm}
\end{figure*}

 {The \emph{spatial bandwidth} of a channel can be defined by the range of effective spatial frequencies included in the channel \cite[p. 392]{Bjornson_BOOK}.}
We will consider the classical $3$\,dB threshold, counting the spatial frequencies with normalized gain above $0.5$. 
 In Fig.~\ref{F_eff_spat}, we plot the effective spatial frequencies with respect to the communication distance $d$ and the angular direction $\theta$. Several important observations can be made:
\begin{itemize}
    \item \textbf{The number of effective spatial frequencies decreases when the distance increases.} From Fig.~\ref{F_eff_spat}(a), we observe that the largest number of effective spatial frequencies is obtained when the user is located at  $d_{\rm B}= {4.5}$\,m. The number of effective spatial frequencies decreases monotonically as the distance increases, becoming just one after $d> {71}$\,m. This has two important implications:
    
    1) Out of the maximum $N=225$ spatial frequencies,   {only less than $50$ spatial frequencies are effective accounting for less than $25 \%$ of the total spatial frequencies}. Hence, despite spherical curvature, the wave components only arrive from a limited range of angles.
        
    2) The number of effective spatial frequencies reduces to one at approximately $d\approx  {71}$\,m, which is around $\frac{d_{\rm FA}}{7}$, where  {$d_{\rm FA}=506.25$\,m} is the Fraunhofer distance. This observation aligns with the $3$\,dB finite beam-depth limit derived in \cite{2024_Kosasih_TWC} for a ULA-like array. Hence, the Fraunhofer distance is a conservative upper limit for the near field. It is only at substantially shorter distances that near-field channels will be much different than  {far}-field channels.
    
    \item \textbf{The number of effective spatial frequencies decreases when the user angle approaches the end-fire directions.}  From Fig.~\ref{F_eff_spat}(b), we observe that the largest number of effective spatial frequencies is obtained when the user is located in the broadside direction of the array ($\theta=\pi/2$). The number of effective spatial frequencies decreases significantly as the angular direction approaches $\pi/2\pm \pi/4$, and then reduces to $1$ as the angular direction approaches $0$ or $\pi$. 
    The reason for this is that the aperture length only seems to be $D_{\rm array} \sin(\theta)$ when viewed from the angle $\theta$, so the effective Fraunhofer distance reduces in non-broadside directions \cite{2024_Kosasih_TWC}. 
    In particular, no near-field phase variations can be observed in the end-fire direction since all antennas  sample the center of the wavefront even it is spherical.
\end{itemize}

In conclusion, a single-path LOS channel can use many spatial frequencies in the near field  {while only a single spatial frequency is occupied in the far field}. However, a half-wavelength-spaced array can only distinguish between $N$ different channel vectors with non-identical spatial frequency content. This could be $N$ far-field channels, $N$ near-field channels, or a mix thereof. Regardless, the spatial DoF of an array is determined by its aperture length. 
 {Although the near-field channel occupies a broader range of spatial frequency resources compared to the far-field channel, which occupies only a single spatial frequency, both utilize the same spatial frequency resources. Therefore, there are no ``new'' spatial dimensions in the near-field case.}
Nevertheless, in a multi-user scenario with random user locations, we typically achieve a higher capacity if some users are in the near field because the channels are then more likely to be distinguishable at the BS
\cite{Bacci2023a}.

\subsection{Multi-User Communication Comparison}

 {The channel characteristics are particularly important when serving multiple users because we want their channel vectors to be as different as possible---ideally mutually orthogonal. We have previously described how $N$ far-field users have orthogonal channels when located in the directions specified by \eqref{eq:directions}.
For each such far-field user, there are additional user locations in the near-field that also give rise to orthogonal channel vectors. However, none of them will be orthogonal to the other $N-1$ far-field users, so every time we add a near-field user, we must remove a far-field user since the total number of spatial frequency resources is $N$.}

 {Orthogonal channels are unlikely to arise in practice; thus, we will analyze the typical communication performance of randomly located users. 
We consider four different cases: 
\begin{itemize}
    \item DFT: Users are deployed according to \eqref{eq:directions} with uniformly spaced directional cosines.
    \item Near field: The users are randomly deployed within the near-field radius of the array $2 D_{\rm array} \leq d \leq d_{\rm FA}/7$.
    \item Far field: The users are randomly deployed within the far-field radius of the array $d \geq d_{\rm FA}$.
    \item Mixed field: Users are randomly deployed within the radius $2 D_{\rm array} \leq d \leq 2d_{\rm FA}$, containing both the near-field and parts of the far-field.
\end{itemize}
To focus on how spatial channel characteristics affect performance, we consider the same signal-to-noise ratio (SNR) for each user, thereby neglecting practical pathloss variations.}

\begin{figure}
\centering
\subfloat[ {Average sum SE over random user locations.}]
{\includegraphics[width=0.49\textwidth]{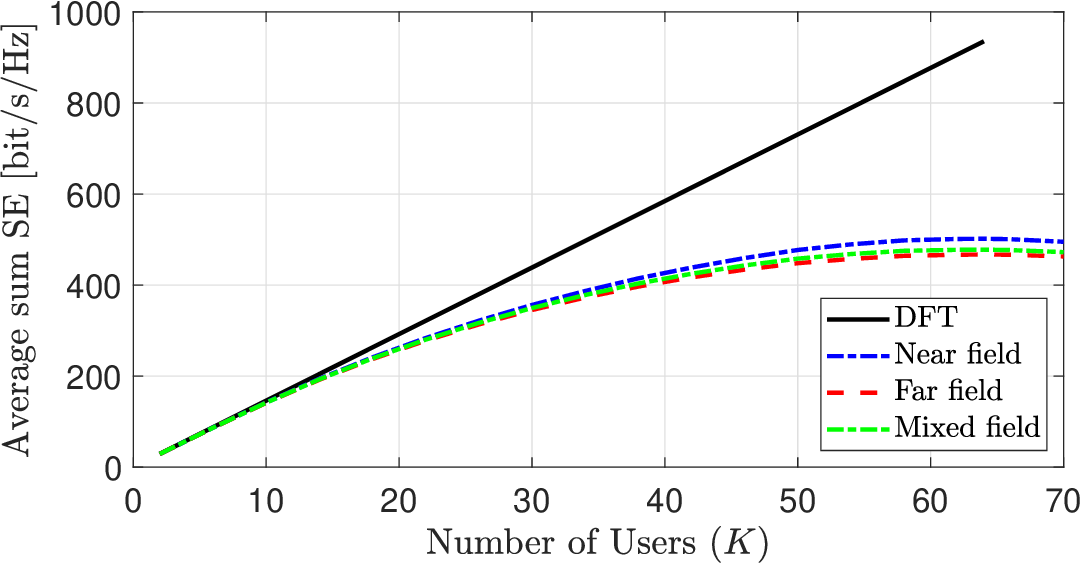}}\hfill
\centering
\subfloat[ {Maximum sum SE over random user locations.}]
{\includegraphics[width=0.49\textwidth]{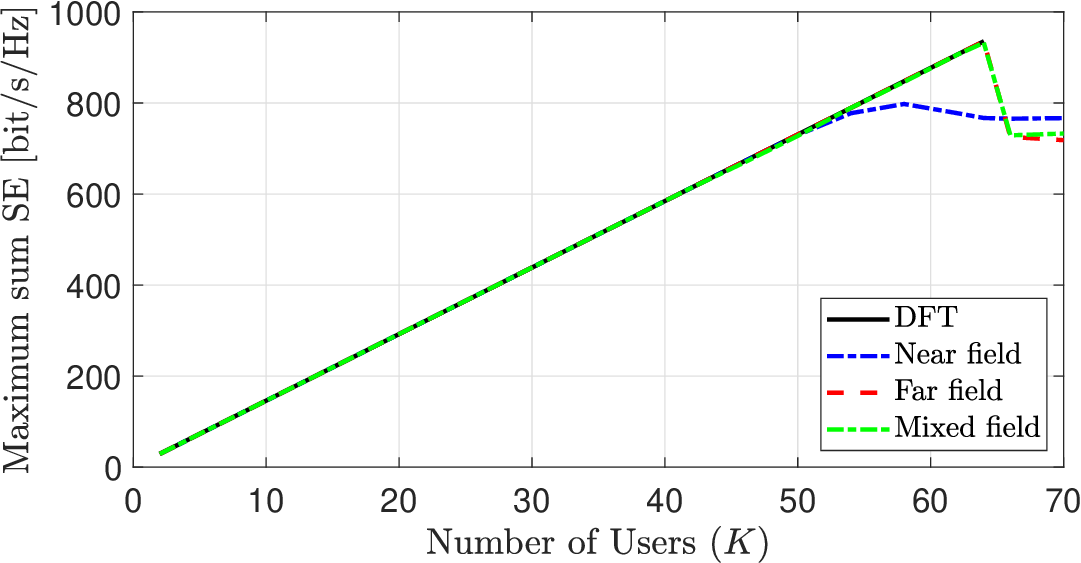}}\hfill
\centering
\subfloat[ {Cumulative distribution function, $N=K=64$.}]
{\includegraphics[width=0.49\textwidth]{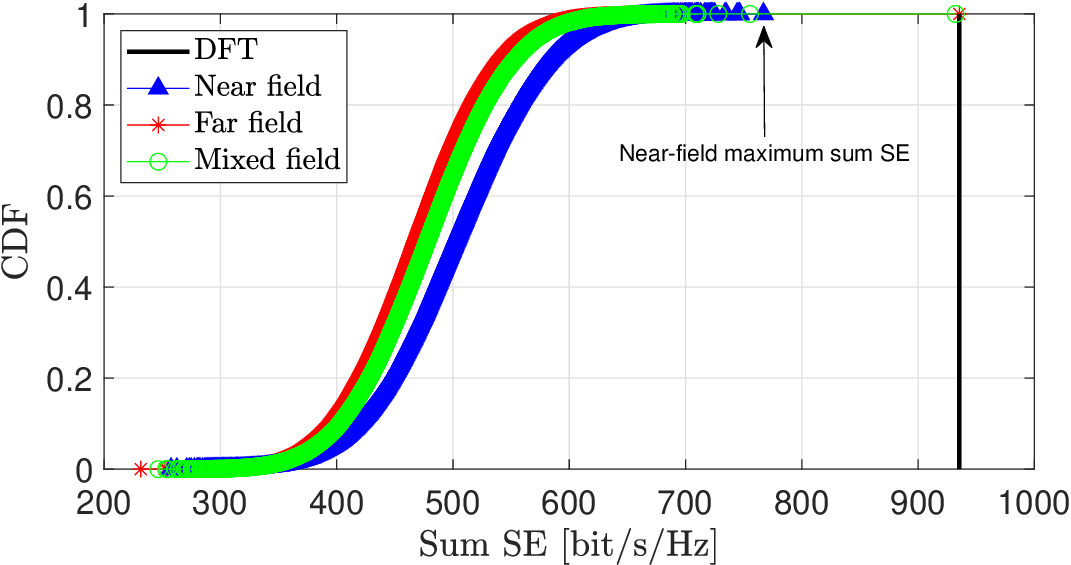}}
\caption{ {Sum SE comparison for multi-user communications with near-field, far-field, and mixed field channels.}}
\label{F_SSE}
\end{figure}

We consider a BS with a ULA composed of $N=64$ half-wavelength-spaced antennas and the carrier frequency is $15$\,GHz. The BS transmits to $K$ users with regularized zero-forcing (RZF) precoding. The downlink sum spectral efficiency (SE) in bit/s/Hz is calculated as in \cite[Ch.~6]{Bjornson_BOOK}. We consider a fixed transmit SNR per user.

In Fig.~\ref{F_SSE}(a), we show the average sum SE across $100\,000$ random user deployments for the above-mentioned different cases. We vary the number of users on the horizontal axis. We notice that the average sum SE in the near-field outperforms the ones in far-field and mixed field. This is thanks to the larger channel variability created when each channel vector depends both on the angle and distance---the BS can distinguish between the users in both domains. However, the ideal DFT case where the user channels are perfectly orthogonal gives a substantially higher SE than all the cases with random user locations.

An additional perspective is provided in Fig.~\ref{F_SSE}(b), where we show the maximum sum SE obtained across the $100\,000$ user deployments. We notice that the maximum sum SEs of the far-field and mixed field cases match with the ideal DFT case for $K \leq N$, meaning that there is at least one random realization among the considered ones that give $N$ (approximately) orthogonal channels.
Interestingly, this is not the case in the near-field, where at most $50$ orthogonal user locations were found. After that point, the sum SE saturates.

The reason behind these results can be seen in Fig.~\ref{F_SSE}(c), which shows the empirical cumulative distribution function (CDF) obtained over the random user deployments with $K=N$ users. We are more likely to obtain a collection of users with decently separable channel vectors when operating in the near-field, but the best-case scenario is not better. We must consider mixed-field or far-field scenarios to find $N$ orthogonal user channels.

 Figs.~\ref{F_SSE}(a) and (b) also show the case where we have $K>N$. We can see that both the average and maximum sum SEs degrade in all the considered scenarios. This highlights the fact that the spatial DoF is $N$, and it is the maximum number of spatial resources that can be used in the near, far, or mixed fields.

\subsection{Spatial Correlation Model for Near-field Channels and Degrees-of-Freedom}

So far, we have considered only LoS channels.  {We will now consider multipath channels, where the non-LoS channel from the ELAA to a single-antenna user can be expressed as the superposition of beams reflected through the objects. We focus on a narrowband channel because the same approach can be applied to study any subcarrier of a wideband system. The channel can be represented by a vector $\vect{h}\in \mathbb{C}^N$ modeled as}
\begin{equation} \label{eq:channel1}
\vect{h} =   \int_{d_{\rm B}}^{\infty}\int_{0}^{\pi} g(\theta,d) \vect{b}(\theta,d)  {d}\theta  {d} d, 
\end{equation}
where the amplitude variations across the antenna array are negligible and $\vect{b}(\theta,d)$ is accurate through the region $d\in[d_{\rm B},\infty)$ and we assume the waves only arrive in front of the array, i.e., $\theta\in[0,\pi]$.  
The \emph{spatial scattering function} $g(\theta,d)$ specifies the gain and phase-shift from each location $(\theta,d)$. 
This function is a generalization of the classical far-field spatial scattering function from \cite{Sayeed2002a} to the near-field channels.
Since scatterers cause small-scale fading, this function is normally modeled as a spatially uncorrelated circularly symmetric complex Gaussian stochastic process. The covariance of $ g(\theta,d)$ and $g(\theta',d')$ is expressed as
\begin{equation} \label{eq:scattering-correlation-model}
\mathbb{E} \left\{ g(\theta,d) g^*\left(\theta',d'\right) \right\} = \beta f(\theta,d) \delta\!\left(\theta-\theta'\right)  \!\delta\!\left(d-d'\right),
\end{equation}
where $(\theta',d')$ represents any arbitrary pair of  angle and distance, $\delta(\cdot)$ denotes the Dirac delta function, $\beta$ denotes the average channel gain, and $f(\theta,d)$ is the normalized \emph{spatial scattering function} \cite{2024_Özlem_Arxiv}. 
It then follows that
\begin{equation} \label{eq:corr-Rayleigh}
\vect{h} \sim \CN(\vect{0},\vect{R}),
\end{equation}
which is a correlated Rayleigh fading channel fully characterized by the spatial correlation matrix
\begin{equation} \label{eq:spatial-correlation}
\vect{R} = \mathbb{E}\{ \vect{h} \vect{h}^{\Htran} \} = \beta \int_{d_{\rm B}}^{\infty}\int_{0}^{\pi} f(\theta,d) \vect{b}(\theta,d) \vect{b}^{\Htran}(\theta,d)  {d} \theta  {d} d,
\end{equation}
where the last equality follows from \eqref{eq:scattering-correlation-model}. 
Hence, even if correlated Rayleigh fading is normally considered for far-field scenarios, it is a general model also applicable to the near field. What differs is the structure of the spatial correlation matrix $\vect{R}$.
By utilizing the structure of the near-field array response vector, we can obtain the $(n,m)$th entry of this matrix as
\begin{align} \label{eq:R-entries}
  \left[\vect{R}\right]_{n,m} = &\beta \int_{d_{\rm B}}^{\infty}\int_{0}^{\pi} f(\theta,d)  e^{\imagunit \frac{2 \pi}{\lambda} \left(  (i_n-i_m) \Delta \cos(\theta)  - \frac{\left(i_n^2-i_m^2\right)  \Delta^2 \left(1-\cos^2(\theta)\right) }{2d} \right)}   {d} \theta  {d} d.
\end{align}
A characteristic feature of spatially correlated channels
is that the eigenvalues of $\vect{R}$ vary drastically \cite{bjornson2024towards}. However, the number of non-negligible eigenvalues, known as the \emph{effective rank} of $\vect{R}$, determines the effective DoF that the ELAA can utilize in the given scattering environment. It is also a limit on the MIMO rank achievable with a multi-antenna user in this scenario.

We will now evaluate the effective rank for a scattering region, where scattering objects are present within a specified angular and distance region. Specifically, the scatters are located at distances $d\in[d_1,d_2]$ for some parameter values $d_1,d_2>0$, and distributed over the angles 
as $\theta\in[\theta_1,\theta_2]$ for some parameter values $\theta_1,\theta_2\in [0,\pi]$.  
To facilitate the computation of the double integral in \eqref{eq:R-entries}, we select $f(\theta,d)$ as
\cite{2024_Özlem_Arxiv}
\begin{align}
f(\theta,d)= \begin{cases}\underbrace{\frac{d_1d_2}{d_2-d_1}\frac{1}{\left(\theta_2-\theta_1\right)}}_{=c} \cdot\frac{1}{d^2}, & d\in[d_1,d_2], \quad \theta\in[\theta_1,\theta_2], \\
0, & \text{otherwise},\end{cases}
\end{align}
where the constant $c$ ensures that $\int_{d_1}^{d_2}\int_{\theta_1}^{\theta_2}f(\theta,d) {d} \theta  {d} d=1$.
Substituting $f(\theta,d)$ into \eqref{eq:spatial-correlation}, we compute all the entries of the spatial correlation matrix $\vect{R}$.

\begin{figure}
    \centering
    \begin{overpic}[width=0.9\textwidth,tics=10]{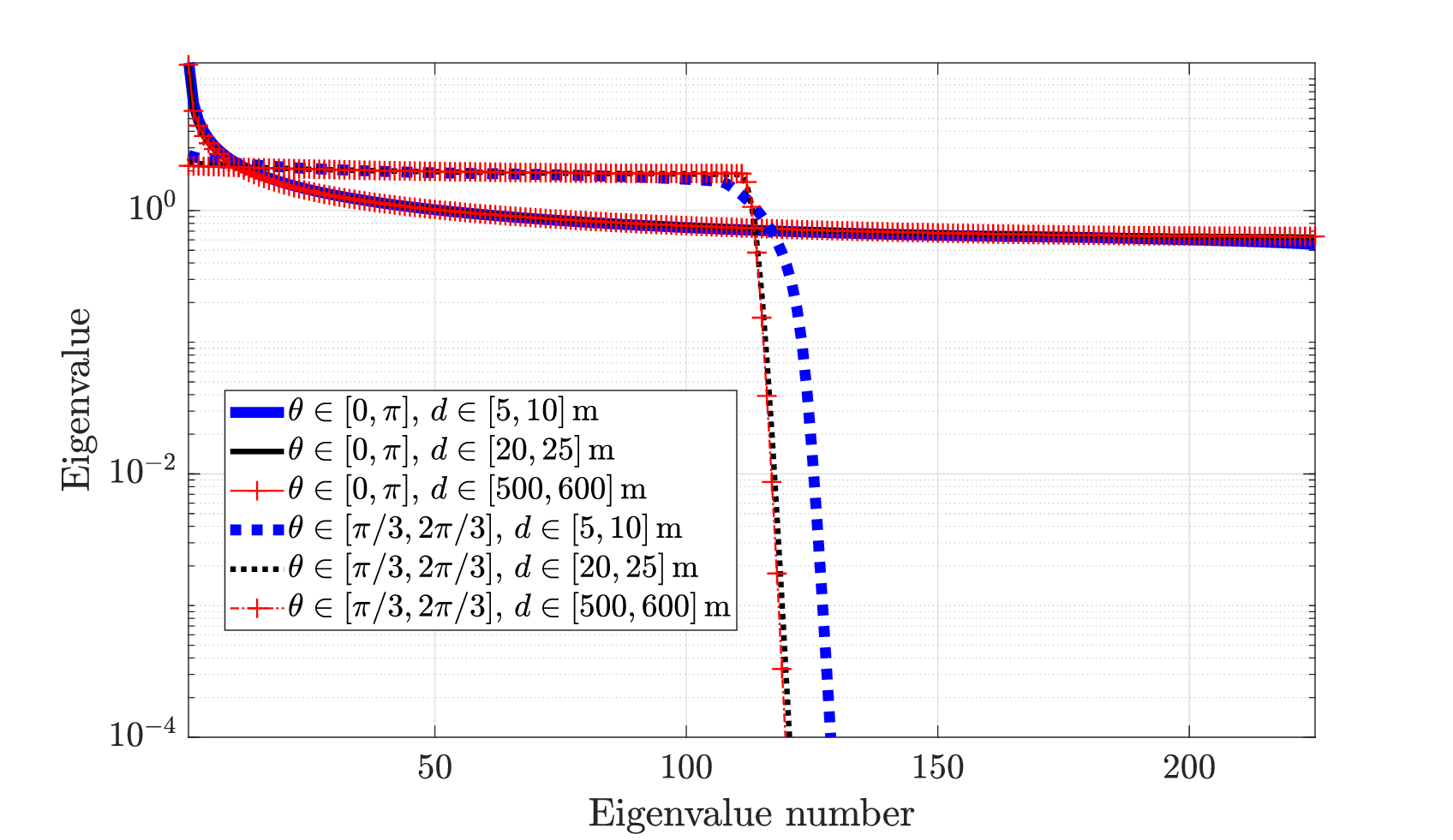}
\end{overpic}
\caption{Eigenvalues of the spatial correlation matrix $\vect{R}$ when scatterers are located in $d\in[d_1,d_2]$ with $\theta\in[\theta_1,\theta_2]$ for $N=225$ and $\Delta=\frac{\lambda}{2}$.}
    \label{fig:eigenvalue-comparison}
       \vspace{-3mm}
\end{figure}

In Fig.~\ref{fig:eigenvalue-comparison}, we plot the sorted eigenvalues of the spatial correlation matrices obtained with several different values of $d_1$, $d_2$, $\theta_1$, and $\theta_2$. We consider $N=225$ antennas with $\Delta=\frac{\lambda}{2}$ and the carrier frequency is $ {15}$\,GHz. The solid lines represent the case where scatterers lie in the whole angular region, i.e., $\theta\in[0,\pi]$. The effective rank is then $N=225$, regardless of the scatterers' distances to the ELAA, demonstrating that the full DoF are achievable in either the near or far field. On the other hand, as the dotted lines show, when $\theta\in {\left[\frac{\pi}{3},\frac{2\pi}{3}\right]}$, there is a reduction in DoF in line with our previous discussion on the reduced number of spatial frequencies. Interestingly, the effective DoF increases from around $ {117}$ to $ {125}$ when the distance of the scatterers is reduced to $d\in  {[5,10]}$\,m. This is expected since there is an energy spread effect in the near field, as previously observed in Fig.~\ref{F_gain_func}. This effect is only visible when the distance is small. In conclusion, the ultimate limit to the DoF is $N=225$, which is attained whenever the whole angular region is covered with scatterers.

%% file: 06_FutureDirections.tex
\section{Research Challenges and Future Directions}
\label{S_Future}

 This section discusses future research challenges and directions within the theme of this paper. We categorize them into three main areas. 
\subsection{Study on Practical Near-Field Channels}

When the array size grows substantially, such as in ELAA, different parts of the array may experience varying multipath propagation environments due to diverse path cluster sets in its vicinity, known as spatial non-stationarity and sometimes modeled by assigning visibility regions to scattering  objects\cite{2023_Bian_TWC,2020_Carvalho_WCMag}. 
This implies that the spatial frequency content is no longer stationary over the array.
This phenomenon comes together with the near-field effect and requires a modification in the channel modeling. The multipath channel can be modeled as a summation of the LoS component of the near-field channel and the non-LoS near-field channels multiplied by an indicator function $\{0,1\}$ depending on the visibility region \cite{2023_Tian_WCLett}. Stochastic near-field channel models of the kind discussed in the previous section must be refined to consider these effects. 
If the non-stationary spatial frequency support of a particular channel can be estimated, this information can be used as prior information for pilot-based channel estimation, perhaps by generalizing the approach in \cite{2024_Özlem_Arxiv}. Once the channel has been estimated, traditional methods can be used for MIMO communications.

Another design aspect that needs to be considered, especially when the inter-antenna spacing is less than $\lambda/2$, is the mutual coupling between the antennas \cite{Chen2018mutual}. This effect can degrade the system's performance, basically by moving energy between spatial frequencies, but will not change the DoF.

Apart from a more realistic channel model, real-world channel measurements are needed to comprehend the propagation behavior in near-field communications and calibrate the models. Recent measurement results are described in \cite{2024_Lu_ComSurv} by listing the essential parameters in various frequency ranges and setups: 1) The first-order statistics, such as channel gain, shadow fading, K-factor, and power-delay profile. 2) The second-order statistics, such as root-mean-squared delay spread. Nevertheless, more comprehensive near-field channel measurements are still needed, especially for the 6G spectrum candidates in the upper mid-band (7-24\,GHz). 3GPP leads studies on this for the upcoming Release-19 of their standards \cite{2023_Lin_Arxiv}.

\subsection{Near-field Beam Training and Beam Tracking}

If the transmitter or receiver is equipped with hybrid analog-digital transceivers, traditional pilot-based estimation methods must be replaced by beam training/tracking.
Far-field beam training aims to identify the LoS direction through hierarchical search on a grid of DFT-like beams that represent such channels.
Near-field beam training is more challenging since near-field LoS channels depend on both angles and distances. 
One option is to send training signals, design a polar-domain codebook with near-field array response vectors and use compressive sensing to find the most matching beam from the codebook \cite{cui2022channel}.
The corresponding complexity is high since the codebook becomes huge.
Alternatively, one can utilize the normalized gain pattern (shown in Fig.~\ref{F_gain_func}) to estimate the center angle. One can then estimate the distance using a reduced-sized
polar-domain codebook \cite{2024_CYou_TWC}. 
Both approaches lead to a massive training overhead when there are many antennas, 
which requires the development of hierarchical near-field beam training methods, possibly using learning-based methods.

When users are served continuously under mobility, the beam training can perform tracking by using the user's last beam to initiate the hierarchical beam search \cite{2023_Zeng_ICC}. This becomes more challenging in the near field since users can simultaneously change their location in both angle and distance. In a system capable of integrated communication and sensing, information obtained regarding the user velocity and direction can be utilized to aid such algorithms. Research in this area is ongoing and since it relies heavily on models, any algorithm must be validated in realistic simulators and measurement campaigns.

\subsection{DoF Optimization and Sparse or Non-Uniform Arrays}

While ELAA offers promising opportunities for massive spatial multiplexing in the enlarged near-field region, deploying or utilizing all the antennas poses significant hardware and processing complexity challenges.
Instead of taking the hybrid beamforming approach, which leads to complex beam tracking issues, one can instead consider large but sparse arrays. Such arrays can benefit from the near-field advantages of ELAA \cite{zhou2024sparse}, but will feature aliasing issues that can potentially be managed by having a non-uniform array geometry.
Alternatively,  one can develop effective antenna selection or utilize movable antennas \cite{2024_Ma_TWC}, to optimize the DoF and the beamforming patterns based on the current number of users and their locations.
The optimal array configuration for a specific scattering environment (and required DoF value) will likely depend on the spatial correlation characteristics. However, a theoretical framework and an algorithmic methodology are yet to be developed.